\documentclass[preprint,12pt]{elsarticle}

\usepackage{amssymb}

\journal{Physics Letters B}

\usepackage{graphicx}  
\usepackage{dcolumn}   
\usepackage{bm}        
\usepackage[breaklinks=true]{hyperref}
\usepackage{amsfonts}   
\usepackage{amssymb}   
\usepackage{amsmath}
\usepackage{verbatim}
\usepackage{url}
\usepackage{epsfig}
\usepackage{natbib}

\begin{document}

\begin{frontmatter}



\title{Semi-Inclusive $\pi^0$ target and beam-target asymmetries from 6 GeV electron scattering with CLAS}

\newcommand*{\ANL}{Argonne National Laboratory, Argonne, Illinois 60439}
\newcommand*{\ANLindex}{1}
\newcommand*{\ASU}{Arizona State University, Tempe, Arizona 85287-1504}
\newcommand*{\ASUindex}{2}
\newcommand*{\CSUDH}{California State University, Dominguez Hills, Carson, CA 90747}
\newcommand*{\CSUDHindex}{3}
\newcommand*{\CANISIUS}{Canisius College, Buffalo, NY}
\newcommand*{\CANISIUSindex}{4}
\newcommand*{\CMU}{Carnegie Mellon University, Pittsburgh, Pennsylvania 15213}
\newcommand*{\CMUindex}{5}
\newcommand*{\CUA}{Catholic University of America, Washington, D.C. 20064}
\newcommand*{\CUAindex}{6}
\newcommand*{\SACLAY}{IRFU, CEA, Universit'e Paris-Saclay, F-91191 Gif-sur-Yvette, France}
\newcommand*{\SACLAYindex}{7}
\newcommand*{\CNU}{Christopher Newport University, Newport News, Virginia 23606}
\newcommand*{\CNUindex}{8}
\newcommand*{\UCONN}{University of Connecticut, Storrs, Connecticut 06269}
\newcommand*{\UCONNindex}{9}
\newcommand*{\FU}{Fairfield University, Fairfield CT 06824}
\newcommand*{\FUindex}{10}
\newcommand*{\FERRARAU}{Universita' di Ferrara , 44121 Ferrara, Italy}
\newcommand*{\FERRARAUindex}{11}
\newcommand*{\FIU}{Florida International University, Miami, Florida 33199}
\newcommand*{\FIUindex}{12}
\newcommand*{\FSU}{Florida State University, Tallahassee, Florida 32306}
\newcommand*{\FSUindex}{13}
\newcommand*{\Genova}{Universit$\grave{a}$ di Genova, 16146 Genova, Italy}
\newcommand*{\Genovaindex}{14}
\newcommand*{\GWUI}{The George Washington University, Washington, DC 20052}
\newcommand*{\GWUIindex}{15}
\newcommand*{\ISU}{Idaho State University, Pocatello, Idaho 83209}
\newcommand*{\ISUindex}{16}
\newcommand*{\INFNFE}{INFN, Sezione di Ferrara, 44100 Ferrara, Italy}
\newcommand*{\INFNFEindex}{17}
\newcommand*{\INFNFR}{INFN, Laboratori Nazionali di Frascati, 00044 Frascati, Italy}
\newcommand*{\INFNFRindex}{18}
\newcommand*{\INFNGE}{INFN, Sezione di Genova, 16146 Genova, Italy}
\newcommand*{\INFNGEindex}{19}
\newcommand*{\INFNRO}{INFN, Sezione di Roma Tor Vergata, 00133 Rome, Italy}
\newcommand*{\INFNROindex}{20}
\newcommand*{\INFNTUR}{INFN, Sezione di Torino, 10125 Torino, Italy}
\newcommand*{\INFNTURindex}{21}
\newcommand*{\ORSAY}{Institut de Physique Nucl\'eaire, CNRS/IN2P3 and Universit\'e Paris Sud, Orsay, France}
\newcommand*{\ORSAYindex}{22}
\newcommand*{\ITEP}{Institute of Theoretical and Experimental Physics, Moscow, 117259, Russia}
\newcommand*{\ITEPindex}{23}
\newcommand*{\JMU}{James Madison University, Harrisonburg, Virginia 22807}
\newcommand*{\JMUindex}{24}
\newcommand*{\KNU}{Kyungpook National University, Daegu 41566, Republic of Korea}
\newcommand*{\KNUindex}{25}
\newcommand*{\MISS}{Mississippi State University, Mississippi State, MS 39762-5167}
\newcommand*{\MISSindex}{26}
\newcommand*{\UNH}{University of New Hampshire, Durham, New Hampshire 03824-3568}
\newcommand*{\UNHindex}{27}
\newcommand*{\NSU}{Norfolk State University, Norfolk, Virginia 23504}
\newcommand*{\NSUindex}{28}
\newcommand*{\OHIOU}{Ohio University, Athens, Ohio  45701}
\newcommand*{\OHIOUindex}{29}
\newcommand*{\ODU}{Old Dominion University, Norfolk, Virginia 23529}
\newcommand*{\ODUindex}{30}
\newcommand*{\URICH}{University of Richmond, Richmond, Virginia 23173}
\newcommand*{\URICHindex}{31}
\newcommand*{\ROMAII}{Universita' di Roma Tor Vergata, 00133 Rome Italy}
\newcommand*{\ROMAIIindex}{32}
\newcommand*{\MSU}{Skobeltsyn Institute of Nuclear Physics, Lomonosov Moscow State University, 119234 Moscow, Russia}
\newcommand*{\MSUindex}{33}
\newcommand*{\SCAROLINA}{University of South Carolina, Columbia, South Carolina 29208}
\newcommand*{\SCAROLINAindex}{34}
\newcommand*{\TEMPLE}{Temple University,  Philadelphia, PA 19122 }
\newcommand*{\TEMPLEindex}{35}
\newcommand*{\JLAB}{Thomas Jefferson National Accelerator Facility, Newport News, Virginia 23606}
\newcommand*{\JLABindex}{36}
\newcommand*{\UTFSM}{Universidad T\'{e}cnica Federico Santa Mar\'{i}a, Casilla 110-V Valpara\'{i}so, Chile}
\newcommand*{\UTFSMindex}{37}
\newcommand*{\EDINBURGH}{Edinburgh University, Edinburgh EH9 3JZ, United Kingdom}
\newcommand*{\EDINBURGHindex}{38}
\newcommand*{\GLASGOW}{University of Glasgow, Glasgow G12 8QQ, United Kingdom}
\newcommand*{\GLASGOWindex}{39}
\newcommand*{\VT}{Virginia Tech, Blacksburg, Virginia   24061-0435}
\newcommand*{\VTindex}{40}
\newcommand*{\VIRGINIA}{University of Virginia, Charlottesville, Virginia 22901}
\newcommand*{\VIRGINIAindex}{41}
\newcommand*{\WM}{College of William and Mary, Williamsburg, Virginia 23187-8795}
\newcommand*{\WMindex}{42}
\newcommand*{\YEREVAN}{Yerevan Physics Institute, 375036 Yerevan, Armenia}
\newcommand*{\YEREVANindex}{43}

\newcommand*{\CORRESP}{Griffioen}
\newcommand*{\NOWJLAB}{Thomas Jefferson National Accelerator Facility, Newport News, Virginia 23606}
\newcommand*{\NOWINFNGE}{INFN, Sezione di Genova, 16146 Genova, Italy}
\newcommand*{\NOWUK}{University of Kentucky, Lexington, Kentucky 40506}
\newcommand*{\NOWISU}{Idaho State University, Pocatello, Idaho 83209}

%
%
\author[toWM]{S ~Jawalkar\fnref{toSANTACLARA}}
\author[toODU]{S. Koirala}
\author[toJLAB]{H.~Avakian}
\author[toWM,toJLAB]{P.~Bosted}
\author[toWM]{K.A.~Griffioen\corref{CORRESP}}
\author[toJLAB]{C.~Keith}
\author[toODU]{S.E.~Kuhn}
\author[toMISS,toFIU]{K.P.~Adhikari}
\author[toFIU]{S.~Adhikari}
\author[toODU]{D.~Adikaram\fnref{toNOWJLAB}}

% ...
\address[toFIU]{\FIU} 
% ...
\address[toMISS]{\MISS} 
% ...
\address[toODU]{\ODU} 
% ...
\address[toJLAB]{\JLAB} 
% ...
\address[toWM]{\WM} 
% ...

 \cortext[CORRESP]{Corresponding author. Email address: griff@wm.edu (K.~Griffioen)}
 \fntext[toSANTACLARA]{Current address: Santa Clara University, Santa Clara, CA 95053}
 \fntext[toNOWJLAB]{Current address: \JLAB}
 \fntext[toNOWINFNGE]{Current address: \Genova}
 \fntext[toGABE]{Current address: Gabes University, 6072-Gabes, Tunisia}
 \fntext[toNOWUK]{Current address: University of Kentucky, Lexington, Kentucky 40506}
 \fntext[SPECTRA]{Current address: Spectral Sciences Inc., 01803 Burlinton, MA}
 \fntext[toNOWISU]{Current address: \ISU}
 \fntext[toORNL]{Current address: Oak Ridge National Laboratory, Oak Ridge, TN 37830}

%
\begin{abstract}
We present precision measurements of the target and beam-target
spin asymmetries from neutral pion electroproduction in deep-inelastic scattering (DIS)
using the CEBAF Large Acceptance
Spectrometer (CLAS) at Jefferson Lab.
We scattered 6-GeV, longitudinally polarized electrons off
longitudinally polarized protons in a cryogenic $^{14}$NH$_3$ target, and extracted
double and single target spin asymmetries for  $ep\rightarrow e^\prime\pi^0X$
in multidimensional bins
in four-momentum transfer ($1.0<Q^2<3.2$ GeV$^2$),
Bjorken-$x$ ($0.12<x<0.48$), hadron energy fraction ($0.4<z<0.7$), transverse pion momentum ($0<P_T<1.0$ GeV), and
azimuthal angle $\phi_h$ between the lepton scattering and hadron production planes.
We extracted asymmetries as a function of both $x$ and $P_T$,  which provide access to
transverse-momentum distributions of longitudinally polarized quarks.
The double spin asymmetries depend weakly on $P_T$.
The  $\sin 2\phi_h$ moments are zero within  uncertainties, which is consistent with
the expected suppression of the Collins fragmentation function.
The observed $\sin\phi_h$ moments
suggest that quark gluon correlations are significant at large $x$.
\end{abstract}


\end{frontmatter}



\newcommand{\xbj}{x}                   
\newcommand{\PT}{P_T}                   


Despite several decades of research, the spin structure of the proton remains 
incompletely understood \cite{Aidala:2012mv}.  
The quark and gluon spins can only partially account for the total proton spin 
of 1/2, leaving the deficit to be found in quark and gluon orbital angular momenta.
The orbital motion of quarks about the proton's
spin axis can be observed in deep-inelastic lepton scattering (DIS) when
a knocked-out quark has momentum transverse to the direction of momentum transfer.  
Although the struck quark acquires transverse momentum in the hadronization process,
there remains enough of a remnant of the original quark orbital motion to probe
quark spin-orbit correlations.  The theoretical motivations and early experiments 
measuring these transverse-momentum distributions (TMDs) have demonstrated that
the theory is sound and the experiments are feasible  \cite{Bacchetta:2016ccz}.
In this Letter we report results of unprecedented accuracy in measurements of spin-azimuthal
asymmetries in neutral pion production in semi-inclusive DIS (SIDIS), 
which provides important information on the quark structure of the proton, 
complementary to that from charged pions.

DIS experiments have mapped the unpolarized structure function 
$f_{1}$ and the polarized structure function $g_{1}$ over a wide range of longitudinal 
momentum fraction $\xbj$ and momentum transfer $Q^2$.  These provide a one-dimensional 
picture of nucleon structure. SIDIS provides access to the 
three-dimensional structure of the nucleon via a new set of structure functions
that depend on the transverse motion of the quarks.  
The scattered lepton and the leading hadron 
are detected in coincidence.  
Eight leading-order ({\it i.e.}\ leading  twist)~\cite{Mulders:1995dh} 
transverse-momentum distributions (TMDs) exist for the different beam and target 
polarizations, which describe the correlations between a quark's transverse momentum
and the spin of the quark or the parent nucleon.  
These correlations manifest themselves in different spin-dependent azimuthal moments of the cross 
section, generated either by correlations in the distribution of quarks or in the fragmentation process, often referred as 
the Sivers ~\cite{Sivers:1989cc} and Collins mechanisms~\cite{Collins:1992kk}, respectively.

For a longitudinally 
polarized nucleon, we have access to two leading-twist TMDs, $g_{1L}$ and $h_{1L}^{\perp}$, 
which respectively describe longitudinally and transversely polarized quarks in a
longitudinally polarized nucleon, and four higher-twist TMDs, $f_{L}^{\perp}$, 
$g_{L}^{\perp}$, $h_{L}$, and $e_{L}$~\cite{Bacchetta:2006tn}
that describe various quark-gluon correlations that vanish as $Q^2\to\infty$.

The HERMES Collaboration made the first observation of a single-spin asymmetry (SSA) in semi-inclusive DIS pion 
electroproduction \cite{Airapetian:1999tv}. This spawned a number 
of additional measurements of SSAs and double spin asymmetries (DSAs) using polarized hydrogen and deuterium targets
\cite{Airapetian:2001eg, Airapetian:2002mf}. 
The target SSAs for proton and deuteron targets published by HERMES 
\cite{Airapetian:2004tw,Airapetian:2004zf, Airapetian:2009ae,Airapetian:2010ds} and
COMPASS ~\cite{Alexakhin:2005iw,Alekseev:2010rw}, provided the first, direct indication
of significant interference terms beyond the simple $s$-wave ($L=0$) picture.  
These asymmetries become larger with increasing $\xbj$,
suggesting that spin-orbit correlations are more relevant for the
valence quarks.  

Measurements of SSAs at
Jefferson Lab (JLab) with longitudinally polarized proton \cite{Avakian:2010ae} and
transversely polarized neutron  
\cite{Qian:2011py,Huang:2011bc,Zhao:2014qvx,Zhang:2013dow} targets suggest that spin-orbit
correlations are significant for certain combinations of 
quark and nucleon spins and transverse momenta. 
Large spin-azimuthal asymmetries were observed at JLab using a longitudinally
polarized beam \cite{Avakian:2003pk,Aghasyan:2011ha} in one case  and a transversely polarized $^3\mathrm{He}$ target in the other 
\cite{Zhao:2015wva}. These results are consistent with the corresponding 
 HERMES~\cite{Airapetian:2005jc} and  COMPASS\cite{Adolph:2014pwc} measurements, which were
interpreted in terms of higher-twist contributions related to quark-gluon correlations.

Previous CLAS measurements \cite{Avakian:2010ae} improved the world data set in two ways:
they showed the first 
hint of a non-zero $\sin 2 \phi_h$ azimuthal moment for charged pions, and
they extracted azimuthal moments in multi-dimensional 
kinematic bins.  COMPASS extended the proton DSAs 
to  low-$x$ \cite{Alekseev:2010ub} using a muon beam
and a polarized NH$_3$ target, and they were able to extract
the dependence on $P_T$, albeit with low statistical accuracy above $x=0.2$.

The world's SSAs and DSAs  are dominated by the charged pion results. 
High statistical accuracy is still needed to study asymmetries as two-dimensional functions of 
$P_T$ and $\xbj$ in order to access the transverse-momentum dependence of different partonic 
distributions, most notably the helicity distribution, $g_1^q$. 
This is true especially for the case of the neutral pion. This paper presents
new results intended to help correct this deficiency.

The electroproduction of neutral pions has several important advantages compared to charged pions: 1)
suppression of higher-twist contributions at large hadron energy fraction $z$ \cite{Afanasev:1996mj},
which are particularly important at JLab energies where small-$z$ events
are contaminated by target fragmentation; 2)
reduction of the background from diffractive $\rho$ decays into pions, which mar
the interpretation of the charged single-pion data;  3) similarity of fragmentation functions for $u$ and
$d$ quarks leading to $\pi^0$, which reduces the dependence
of the DSAs on the fragmentation functions  at large $x$, where valence quarks dominate;
and 4) suppression of spin-dependent
fragmentation for $\pi^0$s, due to the roughly equal magnitude and
opposite sign of the Collins fragmentation functions for up and down quarks
~\cite{Airapetian:2010ds,Alekseev:2010rw,Abe:2005zx,theBABAR:2013yha,Ablikim:2015pta}.
These factors simplify the interpretation of $\pi^0$ SSAs and DSAs. 
Furthermore, neutral pions are straight-forward to identify with little background using
the invariant mass of two detected photons.

The azimuthal  angular dependence  ($\phi_h$) of the asymmetry in the yield for the observed
hadron around the direction of momentum transfer provides our experimental observable.
Longitudinally polarized beams and targets give access to 
longitudinal target SSAs 
and the longitudinal DSAs as a function of $\phi_h$,
4-momentum transfer $Q^2$, Bjorken $x$, transverse hadron momentum $P_T$,
and hadron energy fraction $z$.
These spin asymmetries are defined in the laboratory frame, for which beam and target polarizations
are along  the beam-line (L) or unpolarized (U).
From the $\phi_h$-dependence of these asymmetries (defined on the left-hand side of Eq.\ 1 for SSAs and Eq.\ 2 for
DSAs) we can extract the experimental
azimuthal moments (given on the right-hand side of Eqs.\ 1 and 2) using the $\phi_h$-dependence:
%
\begin{multline}
\left[\frac{1}{P_tf}\right]      \frac{Y^{\downarrow\downarrow}  +  Y^{\uparrow\downarrow} - {Y^{\downarrow\uparrow}  -  Y^{\uparrow\uparrow}}}
 {Y^{\downarrow\downarrow}  +  Y^{\uparrow\downarrow} + {Y^{\downarrow\uparrow}  +  Y^{\uparrow\uparrow}}}=\\
\frac{A_{UL}^{\sin \phi_h} \sin \phi_h + A_{UL}^{\sin 2 \phi_h} \sin 2 \phi_h}{1+A_{UU}^{\cos \phi_h} \cos \phi_h + A_{UU}^{\cos 2 \phi_h} \cos 2 \phi_h}
\end{multline}
and
\begin{multline}
\left[\frac{1}{P_b P_tf}\right]      \frac{Y^{\downarrow\uparrow} +  Y^{\uparrow\downarrow} -  Y^{\uparrow\uparrow} - {Y^{\downarrow\downarrow}  }}
 {Y^{\downarrow\uparrow}  +  Y^{\uparrow\uparrow} + {Y^{\downarrow\downarrow}  +  Y^{\uparrow\downarrow}}}=\\
\frac{A_{LL} + A_{LL}^{\cos \phi_h} \cos \phi_h}{1+A_{UU}^{\cos \phi_h} \cos \phi_h + A_{UU}^{\cos 2 \phi_h} \cos 2 \phi_h}.
\end{multline}
The first (second)  superscript on the yield $Y$ denotes the sign of the beam (target) polarization.
The first (second)  subscript on the azimuthal moment $A$ denotes whether the beam (target) is 
polarized or not.  The superscript on $A$ denotes the azimuthal moment.  No superscript, as in $A_{LL}$, denotes a $\phi_h$-independent
asymmetry.
The angle $\phi_h$ is the hadron azimuthal angle with respect to the lepton plane as defined in the
Trento convention~\cite{Bacchetta:2004jz}.
We normalized the asymmetries using experimentally determined
beam and target polarizations, $P_b$ and $P_t$, respectively, and the dilution factor $f$,
which accounts for the unpolarized material in the target.

In this letter, we present the results for the target SSA $A_{UL}$ and the longitudinal DSA $A_{LL}$  
for $\pi^0$ production in SIDIS using the CLAS detector at JLab \cite{Mecking:2003zu} 
with the addition of a small-angle inner calorimeter (IC) for photons.   
The experiment (eg1-dvcs) took place  from February to October, 2009 
~\cite{Prok:2014ltt,Bosted:2016leu}. We scattered longitudinally polarized 
electrons off a longitudinally polarized solid $^{14}$NH$_3$ target and
collected a total of 30 mC of charge at a beam energy of 5.94 GeV.
We detected scattered electrons and neutral pions in coincidence using CLAS.
The present SIDIS data constitute a subset of our inclusive measurements~\cite{Prok:2014ltt}, and they improve the
older CLAS eg1b $\pi^0$  measurements~\cite{Avakian:2010ae} by an order of magnitude in integrated luminosity.  

We determined the beam polarization (about 85\%) using a M{\o}ller polarimeter \cite{Wagner:1990sn} and 
deduced   the target polarization 
from the product of beam and target polarization (about 65\%) obtained from $ep$ elastic scattering. 
We polarized the protons in $^{14}$NH$_3$ via Dynamic Nuclear Polarization \cite{Crabb:1997cy}.
The CLAS acceptance for scattered electrons $(17^\circ < \theta < 50^\circ)$ was constrained by the IC at small angles and the polarized target 
walls at large angles.

Together, the CLAS electromagnetic calorimeter (EC) and the IC were able to detect
photons from $\pi^0$ decay over a range of angles from $4^\circ$ to  $50^\circ$.
We selected neutral pions by reconstructing the invariant mass of two photons, 
$M_{\gamma \gamma}$~\cite{Kim:2015pkf}.  We analyzed separately three neutral pion topologies,  EC-EC, EC-IC, and IC-IC, 
to take full advantage of the improved energy resolution of the IC and 
the larger angular range of the EC.  Neutral pion mass cuts for EC-EC, EC-IC, and IC-IC were
(0.10,0.17), (0.102,0.17), and (0.105,0.165) GeV, respectively.

Additionally, we applied fiducial cuts to both the EC and IC and removed tracks 
around the edges of the EC where there was a higher negative pion contamination in the electron sample.
We also removed events on the inner edge of the IC (hot blocks close to the 
beam line), as well as blocks on the outer edges of the IC (blocks with incomplete energy reconstruction).
Approximately 4.3M  events survived these cuts.  

We defined our variables using the Trento Convention \cite{Bacchetta:2004jz}, and
selected SIDIS events by imposing kinematic cuts
on the squared 4-momentum transfer ($Q^2 > 1$ GeV$^2$), Bjorken-$x$ ($0.12 < \xbj < 0.48$), the 
target plus virtual photon invariant mass ($W >2$ GeV),
the fractional energy of the $\pi^0$ ($0.40 < z < 0.70$), and the missing mass ($M_x >1.5$ GeV), which
suppressed the contributions from target fragmentation and exclusive events.
We divided the data into  4 bins 
in $\xbj$, 9 bins in $Q^2$,  4 bins in $z$, 6 bins in $\PT$, and 12 bins in $\phi_h$.  
Here, $\phi_h$ is the azimuthal angle 
around the direction of momentum transfer.  Because beam and target polarization lie along the beam direction, all asymmetries
were corrected by a depolarization factor.

We calculated the corresponding SIDIS yields 
by scaling the events by the charge measured with the Faraday Cup in Hall B. 
We scaled the raw asymmetries by the beam and target polarization for $A_{LL}$ and by the 
target polarization for $A_{UL}$. In order to remove contributions from 
the unpolarized part of the $^{14}$NH$_3$ target,  we normalized the raw asymmetries by the
dilution factor (about 3/17), which we calculated using a kinematically dependent model ~\cite{Bosted:2012qc} 
optimized to fit the ratio of SIDIS events \cite{Mineeva:2013hqa} from reference targets. The dilution model takes into 
account the SIDIS cross section per nucleon and an attenuation factor due to final state interactions of the $\pi^0$ in the target.
The relative uncertainty in the dilution factor, due to the determination of 
the length of the frozen target, is 3\%, and the uncertainty from the model dependence is 5\%. Systematic 
uncertainties also resulted from the beam and target polarizations, background 
subtractions, and radiative corrections.
Additionally, we studied the systematic fitting uncertainties for the moment extraction in detail.
The strong dependence of the dilution factor for $\pi^0$s on different kinematic variables
is one of the main sources of systematic uncertainty.  We also estimated via Monte Carlo simulation the
uncertainties on the moment extraction, especially due to
the imprecisely measured $\cos\phi$ and $\cos 2\phi$ dependence in the asymmetry denominators. 

We performed radiative corrections on the data following the theoretical 
developments in Ref.~\cite{Akushevich:2007jc}.
We evaluated the spin-dependent radiative corrections using the Mo-Tsai 
formalism~\cite{Mo:1968cg} in the angle peaking approximation
(photon emission along the incident and scattered electron
directions only) and the equivalent radiator approximation (radiation from
the same nucleus as the hard scattering process is equivalent to
an external radiator of a few percent). We used fits to
the world data on spin-dependent exclusive and inclusive $\pi^0$ electroproduction 
cross sections  and evaluated the radiative
tails for each helicity combination separately
using a Monte-Carlo integration technique. The net effect was
relatively small in most kinematic bins, and is
included in the systematic uncertainty budget.

\begin{figure}
\begin{center}
\includegraphics[width = 8.7cm]{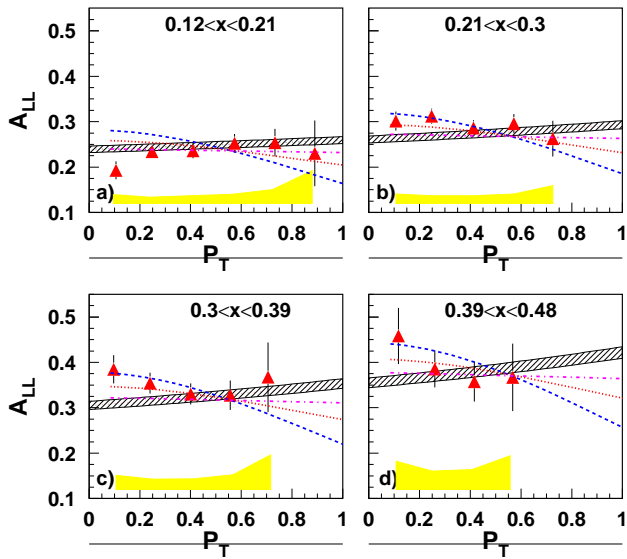}
\caption{The moment $A_{LL}$ versus $\PT$ for $\pi^0$ compared with calculations using the 
quantum  statistical approach to parton distributions~\cite{Bourrely:2005tp,Bourrely:2015kla} (gray bands). The 
dashed, dotted, and dash-dotted
curves are calculations assuming that the $g_1$ to $f_1$ transverse-momentum width ratios are
0.40, 0.68, and 1.0, respectively, using a fixed width for $f_1$ (0.25 GeV$^2$)~\cite{Anselmino:2006yc}.  
The error bars represent the statistical uncertainties, whereas the yellow bands represent the
total experimental systematic uncertainties.
}
\label{fig:all-pt}
\end{center}
\end{figure}

The main goal of this experiment was the extraction of SSAs and DSAs in fine bins in $\xbj$ and transverse hadron momentum $\PT$.
We show here representative results.  
Fig.~\ref{fig:all-pt} shows $A_{LL}$ for $\pi^0$ as a function 
of $P_T$, together with curves 
calculated for our kinematics using different theoretical approaches to parton 
distributions~\cite{Bourrely:2005tp,Bourrely:2015kla}. The general magnitude is predicted well by these calculations, 
while the $P_T$-dependence is less well described.
The dependence of the DSA on $P_T$ indicates that spin orbit correlations may be significant,
and that these dependencies are sensitive to details of the momentum distributions of the polarized quarks.
Because  $A_{LL}$ is related to the ratio of polarized to unpolarized structure functions, this suggest that
transverse momentum is correlated with spin orientation.
Extraction of the underlying quark transverse momentum $k_T$ of the 
helicity distributions, however, will require an established framework for 
TMD extraction from a combination of measurements with unpolarized and polarized targets~\cite{Avakian:2015vha}.

\begin{figure}
\begin{center}
\includegraphics[width = 8.7cm]{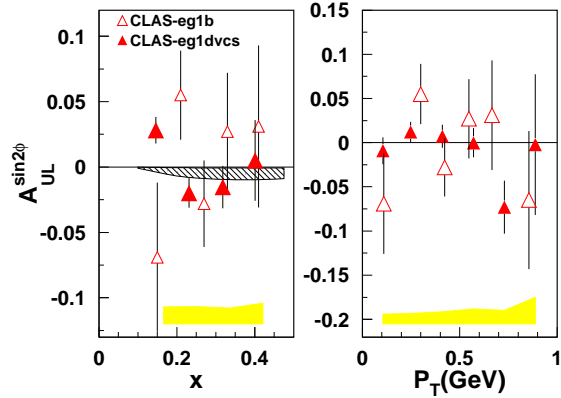}
\caption{The $\sin 2\phi_h$ moments for $A_{UL}$ plotted versus $\xbj$ (left) and $\PT$ (right) compared to 
previous CLAS measurements~\cite{Avakian:2010ae} (which had a lower $z$ threshold of 0.3, no IC, and much lower integrated luminosity)
and theory predictions (gray band) 
~\cite{Bourrely:2005tp,Bourrely:2015kla}.
The error bars represent the statistical uncertainties, whereas the yellow bands represent the
total experimental systematic uncertainties.
}
\label{fig:sin2phi}
\end{center}
\end{figure}

Studies of the Collins fragmentation functions at the $e^+e^-$ machines, BELLE,
\cite{Abe:2005zx,Ogawa:2007zzb,Seidl:2008xc}, 
BABAR~\cite{Garzia:2011vk,theBABAR:2013yha}, and BESIII~\cite{Ablikim:2015pta}, 
indicate that the $\pi^\pm$
Collins fragmentation functions $H_1^\perp$ are large and have
opposite signs for the favored and unfavored cases.  Because fragmentation into $\pi^0$ 
is essentially the average of the $\pi^+$ and $\pi^-$ cases, 
this suggests a significant suppression of the 
Collins fragmentation function for $\pi^0$.  The measured $\sin 2\phi_h$ moment  
of the single target 
spin asymmetry $A_{UL}^{\sin2\phi_h}$, which at leading twist has only a contribution from 
the Collins function coupled to the chiral-odd TMD, $h_{1L}^{\perp}$, is shown in
Fig.~\ref{fig:sin2phi}. This Kotzinian-Mulders SSA~\cite{Kotzinian:1995cz}, 
provides a unique opportunity to check the Collins effect.
Our measurement of  $A_{UL}^{\sin2\phi_h}$ for $\pi^0$ is consistent with zero as expected.

\begin{figure}
\begin{center}
\includegraphics[width = 8.7cm]{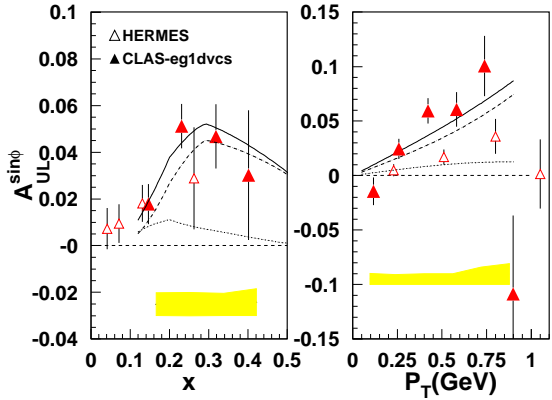}
\caption{The $\sin \phi_h$ moments for $A_{UL}$ vs.\ $x$ (left) and $\PT$ (right). 
The open triangles are the data from HERMES \cite{Airapetian:2002mf},      
and the solid triangles are our new measurements with $z>0.4$. 
The long dashed line is zero for reference.
The short-dashed and dotted lines are twist-3 calculations from
Sivers (larger)  and Collins (smaller) terms~\cite{Mao:2012dk,Lu:2016gtj}, respectively, and the solid line
is the sum of the two. 
The error bars represent the statistical uncertainties, whereas the yellow bands represent the
total experimental systematic uncertainties.
 }
\label{fig:aul-sin-pt}
\end{center}
\end{figure}

A significant $\sin\phi_h$ modulation of the target spin asymmetry has been observed for neutral
pions by the HERMES Collaboration~\cite{Airapetian:2001eg}.
There have been several attempts to describe the $\sin\phi_h$ moment of this asymmetry 
using twist-3 contributions originating from the
unpolarized fragmentation function $D_1$ and the Collins fragmentation function 
$H_1^{\perp}$~\cite{Anselmino:2000mb,Efremov:2002td,Efremov:2001ia,Ma:2002ns}. 
Recently   the effects of the twist-3  TMDs 
$f_L^{\perp}$ and $h_L$ have been calculated in two different 
spectator-diquark models~\cite{Mao:2012dk,Lu:2016gtj}. 
Our data for $A_{UL}^{\sin\phi_h}$ (shown in Fig.~\ref{fig:aul-sin-pt} together with
equaivalent data from HERMES REF at higher beam energies) is plotted versus 
$x$ and $P_T$.  The data suggest that a
Sivers-type contribution coming from the convolution of  $f_L^{\perp}$ and  $D_1$ (dashed 
curve from Ref.~\cite{Lu:2016gtj} in Fig. \ref{fig:aul-sin-pt}) indeed may be dominating 
the $\sin\phi_h$ moment of $A_{UL}$, and quark-gluon correlations are significant for $x>0.2$.

The x-dependence of $A_{UL}$ is consistent with HERMES measurements \cite{Airapetian:2002mf} 
in both magnitude and $x$-dependence.  The increasing $P_T$-dependence is also consistent  with HERMES. 
Precise direct comparisons, however, require taking out the kinematic
factor $\sqrt{2\epsilon(1+\epsilon)}$ from the structure functions, and adding a factor of $ Q$
to account for the higher twist nature of this asymmetry, as
defined in Ref. \cite{Bacchetta:2006tn}. 
Tables with detailed relevant information on double and single target spin asymmetries for  
$ep\rightarrow e^\prime\pi^0X$, extracted for multidimensional bins including $x,z$ and $P_T$-dependences, 
are available at arXiv:1709.10054. 

In summary, kinematic dependencies of single and double spin asymmetries for neutral pions
have been measured in multidimensional bins over a wide kinematic range in $x$
and $P_T$ using CLAS with a polarized proton target.  Measurements of the $P_T$-dependence of  the
double spin asymmetry, performed for the first time for different $\xbj$-bins, indicate the possibility
of different average transverse momenta for quarks aligned or anti-aligned with the nucleon spin.
A non-zero $\sin \phi_h$ target single-spin asymmetry was measured for neutral pions with
high precision, indicating that the target SSA may be generated through the Sivers mechanism.
A small $\sin 2\phi_h$ moment of the target SSA is consistent with expectations
of strong suppression of the Collins effect for neutral pions, due to cancellation of roughly
equal favored and unfavored Collins functions.  
The extent to which higher twist contributes to these extracted moments
at relatively low $Q^2$ constitutes a large part of the upcoming CLAS program with 11 GeV beams.

We thank the accelerator staff, the Physics Division, the Target Group, and the Hall-B staff
 at JLab  for their outstanding efforts that made this experiment possible. 
This work was supported in part by the U.S. Department of Energy
and the National  Science Foundation,
the French Commissariat \`{a} l'Energie Atomique,
the French Centre National de la Recherche Scientifique,
the Italian Istituto Nazionale di Fisica Nucleare, 
the National Research Foundation of Korea,
the United Kingdom's Science and Technology Facilities Council,
and the Southeastern Universities Research Association (SURA), which operates the
Thomas Jefferson National Accelerator Facility for the United States
department of Energy under contract DE-AC05-06OR23177.



\bibliographystyle{elsarticle-num} 
\bibliography{pi0PLB.bib}

\begin{thebibliography}{10}
\expandafter\ifx\csname url\endcsname\relax
  \def\url#1{\texttt{#1}}\fi
\expandafter\ifx\csname urlprefix\endcsname\relax\def\urlprefix{URL }\fi
\expandafter\ifx\csname href\endcsname\relax
  \def\href#1#2{#2} \def\path#1{#1}\fi

\bibitem{Aidala:2012mv}
C.~A. Aidala, S.~D. Bass, D.~Hasch, G.~K. Mallot, {The Spin Structure of the
  Nucleon}, Rev. Mod. Phys. 85 (2013) 655--691.
\newblock \href {http://arxiv.org/abs/1209.2803} {\path{arXiv:1209.2803}},
  \href {http://dx.doi.org/10.1103/RevModPhys.85.655}
  {\path{doi:10.1103/RevModPhys.85.655}}.

\bibitem{Bacchetta:2016ccz}
A.~Bacchetta, {Where do we stand with a 3-D picture of the proton?}, Eur. Phys.
  J. A52~(6) (2016) 163.
\newblock \href {http://dx.doi.org/10.1140/epja/i2016-16163-5}
  {\path{doi:10.1140/epja/i2016-16163-5}}.

\bibitem{Mulders:1995dh}
P.~J. Mulders, R.~D. Tangerman, {The Complete tree level result up to order 1/Q
  for polarized deep inelastic leptoproduction}, Nucl. Phys. B461 (1996)
  197--237, [Erratum:Nucl.Phys.B484,538(1997)].
\newblock \href {http://arxiv.org/abs/hep-ph/9510301}
  {\path{arXiv:hep-ph/9510301}}, \href
  {http://dx.doi.org/10.1016/S0550-3213(96)00648-7,
  10.1016/0550-3213(95)00632-X} {\path{doi:10.1016/S0550-3213(96)00648-7,
  10.1016/0550-3213(95)00632-X}}.

\bibitem{Sivers:1989cc}
D.~W. Sivers, {Single Spin Production Asymmetries from the Hard Scattering of
  Point-Like Constituents}, Phys. Rev. D41 (1990) 83.
\newblock \href {http://dx.doi.org/10.1103/PhysRevD.41.83}
  {\path{doi:10.1103/PhysRevD.41.83}}.

\bibitem{Collins:1992kk}
J.~C. Collins, {Fragmentation of transversely polarized quarks probed in
  transverse momentum distributions}, Nucl. Phys. B396 (1993) 161--182.
\newblock \href {http://arxiv.org/abs/hep-ph/9208213}
  {\path{arXiv:hep-ph/9208213}}, \href
  {http://dx.doi.org/10.1016/0550-3213(93)90262-N}
  {\path{doi:10.1016/0550-3213(93)90262-N}}.

\bibitem{Bacchetta:2006tn}
A.~Bacchetta, M.~Diehl, K.~Goeke, A.~Metz, P.~J. Mulders, M.~Schlegel,
  {Semi-inclusive deep inelastic scattering at small transverse momentum}, JHEP
  02 (2007) 093.
\newblock \href {http://arxiv.org/abs/hep-ph/0611265}
  {\path{arXiv:hep-ph/0611265}}, \href
  {http://dx.doi.org/10.1088/1126-6708/2007/02/093}
  {\path{doi:10.1088/1126-6708/2007/02/093}}.

\bibitem{Airapetian:1999tv}
A.~Airapetian, et~al., {Observation of a single spin azimuthal asymmetry in
  semiinclusive pion electro production}, Phys. Rev. Lett. 84 (2000)
  4047--4051.
\newblock \href {http://arxiv.org/abs/hep-ex/9910062}
  {\path{arXiv:hep-ex/9910062}}, \href
  {http://dx.doi.org/10.1103/PhysRevLett.84.4047}
  {\path{doi:10.1103/PhysRevLett.84.4047}}.

\bibitem{Airapetian:2001eg}
A.~Airapetian, et~al., {Single spin azimuthal asymmetries in electroproduction
  of neutral pions in semiinclusive deep inelastic scattering}, Phys. Rev. D64
  (2001) 097101.
\newblock \href {http://arxiv.org/abs/hep-ex/0104005}
  {\path{arXiv:hep-ex/0104005}}, \href
  {http://dx.doi.org/10.1103/PhysRevD.64.097101}
  {\path{doi:10.1103/PhysRevD.64.097101}}.

\bibitem{Airapetian:2002mf}
A.~Airapetian, et~al., {Measurement of single spin azimuthal asymmetries in
  semiinclusive electroproduction of pions and kaons on a longitudinally
  polarized deuterium target}, Phys. Lett. B562 (2003) 182--192.
\newblock \href {http://arxiv.org/abs/hep-ex/0212039}
  {\path{arXiv:hep-ex/0212039}}, \href
  {http://dx.doi.org/10.1016/S0370-2693(03)00566-5}
  {\path{doi:10.1016/S0370-2693(03)00566-5}}.

\bibitem{Airapetian:2004tw}
A.~Airapetian, et~al., {Single-spin asymmetries in semi-inclusive
  deep-inelastic scattering on a transversely polarized hydrogen target}, Phys.
  Rev. Lett. 94 (2005) 012002.
\newblock \href {http://arxiv.org/abs/hep-ex/0408013}
  {\path{arXiv:hep-ex/0408013}}, \href
  {http://dx.doi.org/10.1103/PhysRevLett.94.012002}
  {\path{doi:10.1103/PhysRevLett.94.012002}}.

\bibitem{Airapetian:2004zf}
A.~Airapetian, et~al., {Quark helicity distributions in the nucleon for up,
  down, and strange quarks from semi-inclusive deep-inelastic scattering},
  Phys. Rev. D71 (2005) 012003.
\newblock \href {http://arxiv.org/abs/hep-ex/0407032}
  {\path{arXiv:hep-ex/0407032}}, \href
  {http://dx.doi.org/10.1103/PhysRevD.71.012003}
  {\path{doi:10.1103/PhysRevD.71.012003}}.

\bibitem{Airapetian:2009ae}
A.~Airapetian, et~al., {Observation of the Naive-T-odd Sivers Effect in
  Deep-Inelastic Scattering}, Phys. Rev. Lett. 103 (2009) 152002.
\newblock \href {http://arxiv.org/abs/0906.3918} {\path{arXiv:0906.3918}},
  \href {http://dx.doi.org/10.1103/PhysRevLett.103.152002}
  {\path{doi:10.1103/PhysRevLett.103.152002}}.

\bibitem{Airapetian:2010ds}
A.~Airapetian, et~al., {Effects of transversity in deep-inelastic scattering by
  polarized protons}, Phys. Lett. B693 (2010) 11--16.
\newblock \href {http://arxiv.org/abs/1006.4221} {\path{arXiv:1006.4221}},
  \href {http://dx.doi.org/10.1016/j.physletb.2010.08.012}
  {\path{doi:10.1016/j.physletb.2010.08.012}}.

\bibitem{Alexakhin:2005iw}
V.~{\relax Yu}. Alexakhin, et~al., {First measurement of the transverse spin
  asymmetries of the deuteron in semi-inclusive deep inelastic scattering},
  Phys. Rev. Lett. 94 (2005) 202002.
\newblock \href {http://arxiv.org/abs/hep-ex/0503002}
  {\path{arXiv:hep-ex/0503002}}, \href
  {http://dx.doi.org/10.1103/PhysRevLett.94.202002}
  {\path{doi:10.1103/PhysRevLett.94.202002}}.

\bibitem{Alekseev:2010rw}
M.~G. Alekseev, et~al., {Measurement of the Collins and Sivers asymmetries on
  transversely polarised protons}, Phys. Lett. B692 (2010) 240--246.
\newblock \href {http://arxiv.org/abs/1005.5609} {\path{arXiv:1005.5609}},
  \href {http://dx.doi.org/10.1016/j.physletb.2010.08.001}
  {\path{doi:10.1016/j.physletb.2010.08.001}}.

\bibitem{Avakian:2010ae}
H.~Avakian, et~al., {Measurement of Single and Double Spin Asymmetries in Deep
  Inelastic Pion Electroproduction with a Longitudinally Polarized Target},
  Phys. Rev. Lett. 105 (2010) 262002.
\newblock \href {http://arxiv.org/abs/1003.4549} {\path{arXiv:1003.4549}},
  \href {http://dx.doi.org/10.1103/PhysRevLett.105.262002}
  {\path{doi:10.1103/PhysRevLett.105.262002}}.

\bibitem{Qian:2011py}
X.~Qian, et~al., {Single Spin Asymmetries in Charged Pion Production from
  Semi-Inclusive Deep Inelastic Scattering on a Transversely Polarized $^3$He
  Target}, Phys. Rev. Lett. 107 (2011) 072003.
\newblock \href {http://arxiv.org/abs/1106.0363} {\path{arXiv:1106.0363}},
  \href {http://dx.doi.org/10.1103/PhysRevLett.107.072003}
  {\path{doi:10.1103/PhysRevLett.107.072003}}.

\bibitem{Huang:2011bc}
J.~Huang, et~al., {Beam-Target Double Spin Asymmetry $A_{LT}$ in Charged Pion
  Production from Deep Inelastic Scattering on a Transversely Polarized He-3
  Target at $1.4<Q^2<2.7$ GeV$^2$}, Phys. Rev. Lett. 108 (2012) 052001.
\newblock \href {http://arxiv.org/abs/1108.0489} {\path{arXiv:1108.0489}},
  \href {http://dx.doi.org/10.1103/PhysRevLett.108.052001}
  {\path{doi:10.1103/PhysRevLett.108.052001}}.

\bibitem{Zhao:2014qvx}
Y.~X. Zhao, et~al., {Single spin asymmetries in charged kaon production from
  semi-inclusive deep inelastic scattering on a transversely polarized $^3He$
  target}, Phys. Rev. C90~(5) (2014) 055201.
\newblock \href {http://arxiv.org/abs/1404.7204} {\path{arXiv:1404.7204}},
  \href {http://dx.doi.org/10.1103/PhysRevC.90.055201}
  {\path{doi:10.1103/PhysRevC.90.055201}}.

\bibitem{Zhang:2013dow}
Y.~Zhang, et~al., {Measurement of pretzelosity asymmetry of charged pion
  production in Semi-Inclusive Deep Inelastic Scattering on a polarized $^3$He
  target}, Phys. Rev. C90~(5) (2014) 055209.
\newblock \href {http://arxiv.org/abs/1312.3047} {\path{arXiv:1312.3047}},
  \href {http://dx.doi.org/10.1103/PhysRevC.90.055209}
  {\path{doi:10.1103/PhysRevC.90.055209}}.

\bibitem{Avakian:2003pk}
H.~Avakian, et~al., {Measurement of beam-spin asymmetries for pi +
  electroproduction above the baryon resonance region}, Phys. Rev. D69 (2004)
  112004.
\newblock \href {http://arxiv.org/abs/hep-ex/0301005}
  {\path{arXiv:hep-ex/0301005}}, \href
  {http://dx.doi.org/10.1103/PhysRevD.69.112004}
  {\path{doi:10.1103/PhysRevD.69.112004}}.

\bibitem{Aghasyan:2011ha}
M.~Aghasyan, et~al., {Precise measurements of beam spin asymmetries in
  semi-inclusive $\pi^0$ production}, Phys. Lett. B704 (2011) 397--402.
\newblock \href {http://arxiv.org/abs/1106.2293} {\path{arXiv:1106.2293}},
  \href {http://dx.doi.org/10.1016/j.physletb.2011.09.044}
  {\path{doi:10.1016/j.physletb.2011.09.044}}.

\bibitem{Zhao:2015wva}
Y.~X. Zhao, et~al., {Double Spin Asymmetries of Inclusive Hadron
  Electroproductions from a Transversely Polarized $^3\rm{He}$ Target}, Phys.
  Rev. C92~(1) (2015) 015207.
\newblock \href {http://arxiv.org/abs/1502.01394} {\path{arXiv:1502.01394}},
  \href {http://dx.doi.org/10.1103/PhysRevC.92.015207}
  {\path{doi:10.1103/PhysRevC.92.015207}}.

\bibitem{Airapetian:2005jc}
A.~Airapetian, et~al., {Subleading-twist effects in single-spin asymmetries in
  semi-inclusive deep-inelastic scattering on a longitudinally polarized
  hydrogen target}, Phys. Lett. B622 (2005) 14--22.
\newblock \href {http://arxiv.org/abs/hep-ex/0505042}
  {\path{arXiv:hep-ex/0505042}}, \href
  {http://dx.doi.org/10.1016/j.physletb.2005.06.067}
  {\path{doi:10.1016/j.physletb.2005.06.067}}.

\bibitem{Adolph:2014pwc}
C.~Adolph, et~al., {Measurement of azimuthal hadron asymmetries in
  semi-inclusive deep inelastic scattering off unpolarised nucleons}, Nucl.
  Phys. B886 (2014) 1046--1077.
\newblock \href {http://arxiv.org/abs/1401.6284} {\path{arXiv:1401.6284}},
  \href {http://dx.doi.org/10.1016/j.nuclphysb.2014.07.019}
  {\path{doi:10.1016/j.nuclphysb.2014.07.019}}.

\bibitem{Alekseev:2010ub}
M.~G. Alekseev, et~al., {Quark helicity distributions from longitudinal spin
  asymmetries in muon-proton and muon-deuteron scattering}, Phys. Lett. B693
  (2010) 227--235.
\newblock \href {http://arxiv.org/abs/1007.4061} {\path{arXiv:1007.4061}},
  \href {http://dx.doi.org/10.1016/j.physletb.2010.08.034}
  {\path{doi:10.1016/j.physletb.2010.08.034}}.

\bibitem{Afanasev:1996mj}
A.~Afanasev, C.~E. Carlson, C.~Wahlquist, {Probing polarized parton
  distributions with meson photoproduction}, Phys. Lett. B398 (1997) 393--399.
\newblock \href {http://arxiv.org/abs/hep-ph/9701215}
  {\path{arXiv:hep-ph/9701215}}, \href
  {http://dx.doi.org/10.1016/S0370-2693(97)00219-0}
  {\path{doi:10.1016/S0370-2693(97)00219-0}}.

\bibitem{Abe:2005zx}
K.~Abe, et~al., {Measurement of azimuthal asymmetries in inclusive production
  of hadron pairs in e+ e- annihilation at Belle}, Phys. Rev. Lett. 96 (2006)
  232002.
\newblock \href {http://arxiv.org/abs/hep-ex/0507063}
  {\path{arXiv:hep-ex/0507063}}, \href
  {http://dx.doi.org/10.1103/PhysRevLett.96.232002}
  {\path{doi:10.1103/PhysRevLett.96.232002}}.

\bibitem{theBABAR:2013yha}
J.~P. Lees, et~al., {Measurement of Collins asymmetries in inclusive production
  of charged pion pairs in $e^+e^-$ annihilation at BABAR}, Phys. Rev. D90~(5)
  (2014) 052003.
\newblock \href {http://arxiv.org/abs/1309.5278} {\path{arXiv:1309.5278}},
  \href {http://dx.doi.org/10.1103/PhysRevD.90.052003}
  {\path{doi:10.1103/PhysRevD.90.052003}}.

\bibitem{Ablikim:2015pta}
M.~Ablikim, et~al., {Measurement of azimuthal asymmetries in inclusive charged
  dipion production in $e^+e^-$ annihilations at $\sqrt{s}$ = 3.65 GeV}, Phys.
  Rev. Lett. 116~(4) (2016) 042001.
\newblock \href {http://arxiv.org/abs/1507.06824} {\path{arXiv:1507.06824}},
  \href {http://dx.doi.org/10.1103/PhysRevLett.116.042001}
  {\path{doi:10.1103/PhysRevLett.116.042001}}.

\bibitem{Bacchetta:2004jz}
A.~Bacchetta, U.~D'Alesio, M.~Diehl, C.~A. Miller, {Single-spin asymmetries:
  The Trento conventions}, Phys. Rev. D70 (2004) 117504.
\newblock \href {http://arxiv.org/abs/hep-ph/0410050}
  {\path{arXiv:hep-ph/0410050}}, \href
  {http://dx.doi.org/10.1103/PhysRevD.70.117504}
  {\path{doi:10.1103/PhysRevD.70.117504}}.

\bibitem{Mecking:2003zu}
B.~A. Mecking, et~al., {The CEBAF Large Acceptance Spectrometer (CLAS)}, Nucl.
  Instrum. Meth. A503 (2003) 513--553.
\newblock \href {http://dx.doi.org/10.1016/S0168-9002(03)01001-5}
  {\path{doi:10.1016/S0168-9002(03)01001-5}}.

\bibitem{Prok:2014ltt}
Y.~Prok, et~al., {Precision measurements of $g_1$ of the proton and the
  deuteron with 6 GeV electrons}, Phys. Rev. C90~(2) (2014) 025212.
\newblock \href {http://arxiv.org/abs/1404.6231} {\path{arXiv:1404.6231}},
  \href {http://dx.doi.org/10.1103/PhysRevC.90.025212}
  {\path{doi:10.1103/PhysRevC.90.025212}}.

\bibitem{Bosted:2016leu}
P.~E. Bosted, et~al., {Target and beam-target spin asymmetries in exclusive
  $\pi^+$ and $\pi^-$ electroproduction with 1.6- to 5.7-GeV electrons}, Phys.
  Rev. C94~(5) (2016) 055201.
\newblock \href {http://arxiv.org/abs/1604.04350} {\path{arXiv:1604.04350}},
  \href {http://dx.doi.org/10.1103/PhysRevC.94.055201}
  {\path{doi:10.1103/PhysRevC.94.055201}}.

\bibitem{Wagner:1990sn}
B.~Wagner, H.~G. Andresen, K.~H. Steffens, W.~Hartmann, W.~Heil, E.~Reichert,
  {A Moller polarimeter for CW and pulsed intermediate-energy electron beams},
  Nucl. Instrum. Meth. A294 (1990) 541--548.
\newblock \href {http://dx.doi.org/10.1016/0168-9002(90)90296-I}
  {\path{doi:10.1016/0168-9002(90)90296-I}}.

\bibitem{Crabb:1997cy}
D.~G. Crabb, W.~Meyer, {Solid polarized targets for nuclear and particle
  physics experiments}, Ann. Rev. Nucl. Part. Sci. 47 (1997) 67--109.
\newblock \href {http://dx.doi.org/10.1146/annurev.nucl.47.1.67}
  {\path{doi:10.1146/annurev.nucl.47.1.67}}.

\bibitem{Kim:2015pkf}
A.~Kim, et~al., {Target and double spin asymmetries of deeply virtual $\pi^0$
  production with a longitudinally polarized proton target and CLAS}, Phys.
  Lett. B768 (2017) 168--173.
\newblock \href {http://arxiv.org/abs/1511.03338} {\path{arXiv:1511.03338}},
  \href {http://dx.doi.org/10.1016/j.physletb.2017.02.032}
  {\path{doi:10.1016/j.physletb.2017.02.032}}.

\bibitem{Bosted:2012qc}
P.~E. Bosted, V.~Mamyan, {Empirical Fit to electron-nucleus scattering},
  Unpublished\href {http://arxiv.org/abs/1203.2262} {\path{arXiv:1203.2262}}.

\bibitem{Mineeva:2013hqa}
T.~Mineeva,
  \href{http://www.jlab.org/Hall-B/general/thesis/Mineeva_thesis.pdf}{{Hadronization
  Studies via $\pi^0$ Electroproduction off D, C, Fe, and Pb}}, Ph.D. thesis,
  Connecticut U. (2013).
\newline\urlprefix\url{http://www.jlab.org/Hall-B/general/thesis/Mineeva_thesis.pdf}

\bibitem{Akushevich:2007jc}
I.~Akushevich, A.~Ilyichev, M.~Osipenko, {Complete lowest order radiative
  corrections to five-fold differential cross-section of hadron
  leptoproduction}, Phys. Lett. B672 (2009) 35--44.
\newblock \href {http://arxiv.org/abs/0711.4789} {\path{arXiv:0711.4789}},
  \href {http://dx.doi.org/10.1016/j.physletb.2008.12.058}
  {\path{doi:10.1016/j.physletb.2008.12.058}}.

\bibitem{Mo:1968cg}
L.~W. Mo, Y.-S. Tsai, {Radiative Corrections to Elastic and Inelastic e p and
  mu p Scattering}, Rev. Mod. Phys. 41 (1969) 205--235.
\newblock \href {http://dx.doi.org/10.1103/RevModPhys.41.205}
  {\path{doi:10.1103/RevModPhys.41.205}}.

\bibitem{Bourrely:2005tp}
C.~Bourrely, J.~Soffer, F.~Buccella, {The Extension to the transverse momentum
  of the statistical parton distributions}, Mod. Phys. Lett. A21 (2006)
  143--150.
\newblock \href {http://arxiv.org/abs/hep-ph/0507328}
  {\path{arXiv:hep-ph/0507328}}, \href
  {http://dx.doi.org/10.1142/S0217732306019244}
  {\path{doi:10.1142/S0217732306019244}}.

\bibitem{Bourrely:2015kla}
C.~Bourrely, J.~Soffer, {New developments in the statistical approach of parton
  distributions: tests and predictions up to LHC energies}, Nucl. Phys. A941
  (2015) 307--334.
\newblock \href {http://arxiv.org/abs/1502.02517} {\path{arXiv:1502.02517}},
  \href {http://dx.doi.org/10.1016/j.nuclphysa.2015.06.018}
  {\path{doi:10.1016/j.nuclphysa.2015.06.018}}.

\bibitem{Avakian:2015vha}
H.~Avakian, H.~Matevosyan, B.~Pasquini, P.~Schweitzer, {Studying the
  information content of TMDs using Monte Carlo generators}, J. Phys. G42
  (2015) 034015.
\newblock \href {http://dx.doi.org/10.1088/0954-3899/42/3/034015}
  {\path{doi:10.1088/0954-3899/42/3/034015}}.

\bibitem{Anselmino:2006yc}
M.~Anselmino, A.~Efremov, A.~Kotzinian, B.~Parsamyan, {Transverse momentum
  dependence of the quark helicity distributions and the Cahn effect in
  double-spin asymmetry A(LL) in Semi Inclusive DIS}, Phys. Rev. D74 (2006)
  074015.
\newblock \href {http://arxiv.org/abs/hep-ph/0608048}
  {\path{arXiv:hep-ph/0608048}}, \href
  {http://dx.doi.org/10.1103/PhysRevD.74.074015}
  {\path{doi:10.1103/PhysRevD.74.074015}}.

\bibitem{Ogawa:2007zzb}
A.~Ogawa, M.~Grosse~Perdekamp, R.-C. Seidl, K.~Hasuko, {Spin dependent
  fragmentation functions analysis at Belle}, AIP Conf. Proc. 915 (2007)
  575--578, [,575(2007)].
\newblock \href {http://dx.doi.org/10.1063/1.2750847}
  {\path{doi:10.1063/1.2750847}}.

\bibitem{Seidl:2008xc}
R.~Seidl, et~al., {Measurement of Azimuthal Asymmetries in Inclusive Production
  of Hadron Pairs in e+e- Annihilation at s**(1/2) = 10.58-GeV}, Phys. Rev. D78
  (2008) 032011, [Erratum: Phys. Rev.D86,039905(2012)].
\newblock \href {http://arxiv.org/abs/0805.2975} {\path{arXiv:0805.2975}},
  \href {http://dx.doi.org/10.1103/PhysRevD.78.032011,
  10.1103/PhysRevD.86.039905} {\path{doi:10.1103/PhysRevD.78.032011,
  10.1103/PhysRevD.86.039905}}.

\bibitem{Garzia:2011vk}
I.~Garzia, {Measurement of Collins asymmetries in the inclusive production of
  pion pairs in electron-positron collisions at BaBar}, Nuovo Cim. C034N06
  (2011) 49--51.
\newblock \href {http://dx.doi.org/10.1393/ncc/i2011-11034-5}
  {\path{doi:10.1393/ncc/i2011-11034-5}}.

\bibitem{Kotzinian:1995cz}
A.~M. Kotzinian, P.~J. Mulders, {Longitudinal quark polarization in
  transversely polarized nucleons}, Phys. Rev. D54 (1996) 1229--1232.
\newblock \href {http://arxiv.org/abs/hep-ph/9511420}
  {\path{arXiv:hep-ph/9511420}}, \href
  {http://dx.doi.org/10.1103/PhysRevD.54.1229}
  {\path{doi:10.1103/PhysRevD.54.1229}}.

\bibitem{Anselmino:2000mb}
M.~Anselmino, F.~Murgia, {Spin effects in the fragmentation of a transversely
  polarized quark}, Phys. Lett. B483 (2000) 74--86.
\newblock \href {http://arxiv.org/abs/hep-ph/0002120}
  {\path{arXiv:hep-ph/0002120}}, \href
  {http://dx.doi.org/10.1016/S0370-2693(00)00519-0}
  {\path{doi:10.1016/S0370-2693(00)00519-0}}.

\bibitem{Efremov:2002td}
A.~V. Efremov, K.~Goeke, P.~Schweitzer, {Azimuthal asymmetry in
  electroproduction of neutral pions in semiinclusive DIS}, Phys. Lett. B522
  (2001) 37--48, [Erratum: Phys. Lett.B544,389(2002)].
\newblock \href {http://arxiv.org/abs/hep-ph/0108213}
  {\path{arXiv:hep-ph/0108213}}, \href
  {http://dx.doi.org/10.1016/S0370-2693(01)01258-8,
  10.1016/S0370-2693(02)02518-2} {\path{doi:10.1016/S0370-2693(01)01258-8,
  10.1016/S0370-2693(02)02518-2}}.

\bibitem{Efremov:2001ia}
A.~V. Efremov, K.~Goeke, P.~Schweitzer, {Predictions for azimuthal asymmetries
  in pion and kaon production in SIDIS off a longitudinally polarized deuterium
  target at HERMES}, Eur. Phys. J. C24 (2002) 407--412.
\newblock \href {http://arxiv.org/abs/hep-ph/0112166}
  {\path{arXiv:hep-ph/0112166}}, \href
  {http://dx.doi.org/10.1007/s100520200918} {\path{doi:10.1007/s100520200918}}.

\bibitem{Ma:2002ns}
B.-Q. Ma, I.~Schmidt, J.-J. Yang, {Reanalysis of azimuthal spin asymmetries of
  meson electroproduction}, Phys. Rev. D66 (2002) 094001.
\newblock \href {http://arxiv.org/abs/hep-ph/0209114}
  {\path{arXiv:hep-ph/0209114}}, \href
  {http://dx.doi.org/10.1103/PhysRevD.66.094001}
  {\path{doi:10.1103/PhysRevD.66.094001}}.

\bibitem{Mao:2012dk}
W.~Mao, Z.~Lu, {Beam single spin asymmetry of neutral pion production in
  semi-inclusive deep inelastic scattering}, Phys. Rev. D87~(1) (2013) 014012.
\newblock \href {http://arxiv.org/abs/1210.4790} {\path{arXiv:1210.4790}},
  \href {http://dx.doi.org/10.1103/PhysRevD.87.014012}
  {\path{doi:10.1103/PhysRevD.87.014012}}.

\bibitem{Lu:2016gtj}
Z.~Lu, W.~Mao, {Single-Spin Asymmetries $A_{UL}^{sin \phi h}$ in Semi-Inclusive
  Pions Production}, Int. J. Mod. Phys. Conf. Ser. 40 (2016) 1660045.
\newblock \href {http://dx.doi.org/10.1142/S2010194516600454}
  {\path{doi:10.1142/S2010194516600454}}.

\end{thebibliography}





\eject

%
%


\addtolength{\oddsidemargin}{-3cm}
\addtolength{\evensidemargin}{-3cm}
\addtolength{\topmargin}{-3cm}
\addtolength{\textheight}{3cm}
\scriptsize

\section{Table 1: $A_{LL}^{\rm const} vs.\ P_T$}
\begin{verbatim}
%% CLAS eg1-dvcs SIDIS azimuthal asymmetries for neutral pions
%% The table quotes the average value within a bin.  
%% P_T is the momentum of the pi0 transverse to the momentum transfer direction (in GeV).
%% x is the Bjorken scaling variable.
%% A_LL is the measured double spin asymmetry for longitudinally polarized electrons and protons.
%% Delta_stat is the statistical uncertainty on A_LL. 
%% Delta_sys is the systematic uncertainty on A_LL.
%% y is the fractional energy transfer.
%% z is the fractional energy of the pi0.
%% Q^2 is the 4-momentum transfer squared.
%% depol is the depolarization factor, which accounts for the mismatch between the beam and momentum transfer directions.
%% dilut is the applied dilution factor.

  P_T-bin#  P_T       x      A_LL  Delta stat Delta_sys   y        z      Q^2     depol     dilut
      1    0.106    0.168    0.187    0.020    0.015    0.748    0.534    1.394    0.882    0.185
      2    0.248    0.166    0.231    0.014    0.016    0.752    0.528    1.382    0.885    0.181
      3    0.410    0.165    0.239    0.017    0.019    0.757    0.523    1.379    0.888    0.175
      4    0.574    0.168    0.253    0.021    0.023    0.745    0.521    1.384    0.878    0.166
      5    0.733    0.166    0.254    0.031    0.020    0.754    0.521    1.390    0.887    0.154
      6    0.890    0.160    0.219    0.073    0.037    0.779    0.495    1.388    0.909    0.140
:::::::::
      1    0.106    0.250    0.309    0.021    0.019    0.717    0.535    1.990    0.871    0.196
      2    0.247    0.251    0.316    0.017    0.020    0.704    0.531    1.963    0.858    0.193
      3    0.413    0.252    0.287    0.018    0.022    0.688    0.510    1.924    0.842    0.186
      4    0.572    0.251    0.297    0.021    0.023    0.697    0.503    1.942    0.853    0.176
      5    0.725    0.247    0.263    0.040    0.018    0.737    0.500    2.024    0.892    0.164
      6    0.879    0.240    0.038    0.499    0.037    0.786    0.463    2.101    0.936    0.149
:::::::::
      1    0.106    0.338    0.385    0.032    0.022    0.698    0.527    2.622    0.875    0.206
      2    0.250    0.339    0.358    0.023    0.021    0.669    0.517    2.520    0.845    0.203
      3    0.411    0.338    0.327    0.024    0.025    0.658    0.493    2.475    0.835    0.195
      4    0.566    0.336    0.313    0.033    0.021    0.698    0.483    2.609    0.877    0.185
      5    0.716    0.331    0.348    0.077    0.029    0.759    0.476    2.800    0.935    0.173
:::::::::
      1    0.107    0.421    0.460    0.063    0.031    0.678    0.512    3.167    0.880    0.214
      2    0.250    0.422    0.376    0.041    0.024    0.653    0.496    3.062    0.855    0.210
      3    0.406    0.422    0.352    0.044    0.025    0.658    0.471    3.085    0.862    0.202
      4    0.557    0.416    0.377    0.075    0.030    0.714    0.460    3.304    0.917    0.192
\end{verbatim}
\eject

\section{Table 2: $A_{LL}^{\rm const} vs.\ x$}
\begin{verbatim}
%% CLAS eg1-dvcs SIDIS azimuthal asymmetries for neutral pions
%% The table quotes the average value within a bin.
%% P_T is the momentum of the pi0 transverse to the momentum transfer direction (in GeV).
%% x is the Bjorken scaling variable.
%% A_LL is the measured double spin asymmetry for longitudinally polarized electrons and protons.
%% Delta_stat is the statistical uncertainty on A_LL.
%% Delta_sys is the systematic uncertainty on A_LL.
%% y is the fractional energy transfer.
%% z is the fractional energy of the pi0.
%% Q^2 is the 4-momentum transfer squared.
%% depol is the depolarization factor, which accounts for the mismatch between the beam and momentum transfer directions.
%% dilut is the applied dilution factor.


   x-bin#  P_T       x      A_LL  \Delta A_stat Delta_sys   y      z       Q^2     depol    dilut
      1    0.106    0.168    0.187    0.020    0.015    0.748    0.534    1.394    0.882    0.185
      2    0.106    0.250    0.309    0.021    0.019    0.717    0.535    1.990    0.871    0.196
      3    0.106    0.338    0.385    0.032    0.022    0.698    0.527    2.622    0.875    0.206
      4    0.107    0.421    0.460    0.063    0.031    0.678    0.512    3.167    0.880    0.214
:::::::::
      1    0.248    0.166    0.231    0.014    0.016    0.752    0.528    1.382    0.885    0.181
      2    0.247    0.251    0.316    0.017    0.020    0.704    0.531    1.963    0.858    0.193
      3    0.250    0.339    0.358    0.023    0.021    0.669    0.517    2.520    0.845    0.203
      4    0.250    0.422    0.376    0.041    0.024    0.653    0.496    3.062    0.855    0.210
:::::::::
      1    0.410    0.165    0.239    0.017    0.019    0.757    0.523    1.379    0.888    0.175
      2    0.413    0.252    0.287    0.018    0.022    0.688    0.510    1.924    0.842    0.186
      3    0.411    0.338    0.327    0.024    0.025    0.658    0.493    2.475    0.835    0.195
      4    0.406    0.422    0.352    0.044    0.025    0.658    0.471    3.085    0.862    0.202
:::::::::
      1    0.574    0.168    0.253    0.021    0.023    0.745    0.521    1.384    0.878    0.166
      2    0.572    0.251    0.297    0.021    0.023    0.697    0.503    1.942    0.853    0.176
      3    0.566    0.336    0.313    0.033    0.021    0.698    0.483    2.609    0.877    0.185
      4    0.557    0.416    0.377    0.075    0.030    0.714    0.460    3.304    0.917    0.192
:::::::::
      1    0.733    0.166    0.254    0.031    0.020    0.754    0.521    1.390    0.887    0.154
      2    0.725    0.247    0.263    0.040    0.018    0.737    0.500    2.024    0.892    0.164
      3    0.716    0.331    0.348    0.077    0.029    0.759    0.476    2.800    0.935    0.173
:::::::::
      1    0.890    0.160    0.219    0.073    0.037    0.779    0.495    1.388    0.909    0.140
      2    0.879    0.240    0.038    0.499    0.037    0.786    0.463    2.101    0.936    0.149
\end{verbatim}
\eject

\section{Table 3: $A_{LL}^{\cos\phi_h}  vs.\ P_T$}
\begin{verbatim}
%% CLAS eg1-dvcs SIDIS azimuthal asymmetries for neutral pions
%% The table quotes the average value within a bin.
%% P_T is the momentum of the pi0 transverse to the momentum transfer direction (in GeV).
%% x is the Bjorken scaling variable.
%% A_LL^cos is the measured cosine-phi moment of the double spin asymmetry for longitudinally polarized electrons and protons.
%% Delta_stat is the statistical uncertainty on the moment.
%% Delta_sys is the systematic uncertainty on the moment.
%% y is the fractional energy transfer.
%% z is the fractional energy of the pi0.
%% Q^2 is the 4-momentum transfer squared.
%% depol is the depolarization factor, which accounts for the mismatch between the beam and momentum transfer directions.
%% dilut is the applied dilution factor.


  P_T-bin#  P_T       x    A_LL^cos  Delta_stat Delta_sys  y      z       Q^2     depol     dilut
      1    0.106    0.168   -0.002    0.030    0.049    0.748    0.534    1.394    0.882    0.185
      2    0.248    0.166    0.011    0.024    0.055    0.752    0.528    1.382    0.885    0.181
      3    0.410    0.165   -0.025    0.025    0.055    0.757    0.523    1.379    0.888    0.175
      4    0.574    0.168   -0.022    0.026    0.062    0.745    0.521    1.384    0.878    0.166
      5    0.733    0.166   -0.038    0.040    0.065    0.754    0.521    1.390    0.887    0.154
      6    0.890    0.160   -0.054    0.107    0.034    0.779    0.495    1.388    0.909    0.140
:::::::::
      1    0.106    0.250   -0.037    0.030    0.074    0.717    0.535    1.990    0.871    0.196
      2    0.247    0.251   -0.038    0.025    0.075    0.704    0.531    1.963    0.858    0.193
      3    0.413    0.252   -0.022    0.024    0.071    0.688    0.510    1.924    0.842    0.186
      4    0.572    0.251    0.006    0.027    0.071    0.697    0.503    1.942    0.853    0.176
      5    0.725    0.247   -0.058    0.052    0.051    0.737    0.500    2.024    0.892    0.164
      6    0.879    0.240   -0.344    0.241    0.043    0.786    0.463    2.101    0.936    0.149
:::::::::
      1    0.106    0.338   -0.002    0.044    0.091    0.698    0.527    2.622    0.875    0.206
      2    0.250    0.339   -0.006    0.031    0.090    0.669    0.517    2.520    0.845    0.203
      3    0.411    0.338   -0.050    0.031    0.085    0.658    0.493    2.475    0.835    0.195
      4    0.566    0.336   -0.024    0.041    0.068    0.698    0.483    2.609    0.877    0.185
      5    0.716    0.331   -0.161    0.099    0.072    0.759    0.476    2.800    0.935    0.173
:::::::::
      1    0.107    0.421    0.039    0.083    0.105    0.678    0.512    3.167    0.880    0.214
      2    0.250    0.422   -0.028    0.052    0.103    0.653    0.496    3.062    0.855    0.210
      3    0.406    0.422    0.055    0.057    0.096    0.658    0.471    3.085    0.862    0.202
      4    0.557    0.416   -0.070    0.094    0.106    0.714    0.460    3.304    0.917    0.192
\end{verbatim}
\eject

\section{Table 4: $A_{LL}^{\cos\phi_h}  vs.\ x$}
\begin{verbatim}
%% CLAS eg1-dvcs SIDIS azimuthal asymmetries for neutral pions
%% The table quotes the average value within a bin.
%% P_T is the momentum of the pi0 transverse to the momentum transfer direction (in GeV).
%% x is the Bjorken scaling variable.
%% A_LL^cos is the cosine-phi moment of the measured double spin asymmetry for longitudinally polarized electrons and protons.
%% Delta_stat is the statistical uncertainty on the moment.
%% Delta_sys is the systematic uncertainty on the moment.
%% y is the fractional energy transfer.
%% z is the fractional energy of the pi0.
%% Q^2 is the 4-momentum transfer squared.
%% depol is the depolarization factor, which accounts for the mismatch between the beam and momentum transfer directions.
%% dilut is the applied dilution factor.


   x-bin#  P_T       x    A_LL^cos  Delta_stat Delta_sys  y        z      Q^2     depol     dilut
      1    0.106    0.168   -0.002    0.030    0.049    0.748    0.534    1.394    0.882    0.185
      2    0.106    0.250   -0.037    0.030    0.074    0.717    0.535    1.990    0.871    0.196
      3    0.106    0.338   -0.002    0.044    0.091    0.698    0.527    2.622    0.875    0.206
      4    0.107    0.421    0.039    0.083    0.105    0.678    0.512    3.167    0.880    0.214
:::::::::
      1    0.248    0.166    0.011    0.024    0.055    0.752    0.528    1.382    0.885    0.181
      2    0.247    0.251   -0.038    0.025    0.075    0.704    0.531    1.963    0.858    0.193
      3    0.250    0.339   -0.006    0.031    0.090    0.669    0.517    2.520    0.845    0.203
      4    0.250    0.422   -0.028    0.052    0.103    0.653    0.496    3.062    0.855    0.210
:::::::::
      1    0.410    0.165   -0.025    0.025    0.055    0.757    0.523    1.379    0.888    0.175
      2    0.413    0.252   -0.022    0.024    0.071    0.688    0.510    1.924    0.842    0.186
      3    0.411    0.338   -0.050    0.031    0.085    0.658    0.493    2.475    0.835    0.195
      4    0.406    0.422    0.055    0.057    0.096    0.658    0.471    3.085    0.862    0.202
:::::::::
      1    0.574    0.168   -0.022    0.026    0.062    0.745    0.521    1.384    0.878    0.166
      2    0.572    0.251    0.006    0.027    0.071    0.697    0.503    1.942    0.853    0.176
      3    0.566    0.336   -0.024    0.041    0.068    0.698    0.483    2.609    0.877    0.185
      4    0.557    0.416   -0.070    0.094    0.106    0.714    0.460    3.304    0.917    0.192
:::::::::
      1    0.733    0.166   -0.038    0.040    0.065    0.754    0.521    1.390    0.887    0.154
      2    0.725    0.247   -0.058    0.052    0.051    0.737    0.500    2.024    0.892    0.164
      3    0.716    0.331   -0.161    0.099    0.072    0.759    0.476    2.800    0.935    0.173
:::::::::
      1    0.890    0.160   -0.054    0.107    0.034    0.779    0.495    1.388    0.909    0.140
      2    0.879    0.240   -0.344    0.241    0.043    0.786    0.463    2.101    0.936    0.149
\end{verbatim}
\eject

\section{Table 5: $A_{UL}^{\sin\phi_h}  vs.\ P_T$}
\begin{verbatim}
%% CLAS eg1-dvcs SIDIS azimuthal asymmetries for neutral pions
%% The table quotes the average value within a bin.
%% P_T is the momentum of the pi0 transverse to the momentum transfer direction (in GeV).
%% x is the Bjorken scaling variable.
%% A_ULsin is the measured sine-phi moment of the target spin asymmetry for longitudinally polarized protons.
%% Delta_stat is the statistical uncertainty on the moment.
%% Delta_sys is the systematic uncertainty on the moment.
%% y is the fractional energy transfer.
%% z is the fractional energy of the pi0.
%% Q^2 is the 4-momentum transfer squared.
%% depol is the depolarization factor, which accounts for the mismatch between the beam and momentum transfer directions.
%% dilut is the applied dilution factor.


  P_T-bin#  P_T       x     A_ULsin  Delta stat Delta_sys   y       z     Q^2     depol     dilut
      1    0.106    0.168   -0.018    0.023    0.011    0.748    0.534    1.394    0.882    0.185
      2    0.248    0.166    0.006    0.016    0.011    0.752    0.528    1.382    0.885    0.181
      3    0.410    0.165    0.029    0.020    0.011    0.757    0.523    1.379    0.888    0.175
      4    0.574    0.168    0.032    0.029    0.010    0.745    0.521    1.384    0.878    0.166
      5    0.733    0.166    0.164    0.041    0.018    0.754    0.521    1.390    0.887    0.154
      6    0.890    0.160   -0.135    0.084    0.019    0.779    0.495    1.388    0.909    0.140
:::::::::
      1    0.106    0.250   -0.038    0.026    0.011    0.717    0.535    1.990    0.871    0.196
      2    0.247    0.251    0.056    0.020    0.011    0.704    0.531    1.963    0.858    0.193
      3    0.413    0.252    0.138    0.025    0.013    0.688    0.510    1.924    0.842    0.186
      4    0.572    0.251    0.116    0.030    0.012    0.697    0.503    1.942    0.853    0.176
      5    0.725    0.247    0.027    0.053    0.019    0.737    0.500    2.024    0.892    0.164
      6    0.879    0.240   -0.030    0.165    0.023    0.786    0.463    2.101    0.936    0.149
:::::::::
      1    0.106    0.338    0.004    0.038    0.011    0.698    0.527    2.622    0.875    0.206
      2    0.250    0.339    0.069    0.031    0.011    0.669    0.517    2.520    0.845    0.203
      3    0.411    0.338    0.074    0.034    0.013    0.658    0.493    2.475    0.835    0.195
      4    0.566    0.336    0.075    0.046    0.014    0.698    0.483    2.609    0.877    0.185
      5    0.716    0.331    0.108    0.109    0.013    0.759    0.476    2.800    0.935    0.173
:::::::::
      1    0.107    0.421    0.103    0.077    0.018    0.678    0.512    3.167    0.880    0.214
      2    0.250    0.422   -0.009    0.060    0.012    0.653    0.496    3.062    0.855    0.210
      3    0.406    0.422    0.056    0.066    0.018    0.658    0.471    3.085    0.862    0.202
      4    0.557    0.416    0.053    0.109    0.017    0.714    0.460    3.304    0.917    0.192
\end{verbatim}
\eject

\section{Table 6: $A_{UL}^{\sin\phi_h}  vs.\ x$}
\begin{verbatim}
%% CLAS eg1-dvcs SIDIS azimuthal asymmetries for neutral pions
%% The table quotes the average value within a bin.
%% P_T is the momentum of the pi0 transverse to the momentum transfer direction (in GeV).
%% x is the Bjorken scaling variable.
%% A_ULsin is the measured sine-phi moment of the target spin asymmetry for longitudinally polarized protons.
%% Delta_stat is the statistical uncertainty on the moment.
%% Delta_sys is the systematic uncertainty on the moment.
%% y is the fractional energy transfer.
%% z is the fractional energy of the pi0.
%% Q^2 is the 4-momentum transfer squared.
%% depol is the depolarization factor, which accounts for the mismatch between the beam and momentum transfer directions.
%% dilut is the applied dilution factor.

  x-bin#  P_T       x      A_ULsin  Delta stat Delta_sys  y        z      Q^2     depol     dilut
      1    0.106    0.168   -0.018    0.023    0.011    0.748    0.534    1.394    0.882    0.185
      2    0.106    0.250   -0.038    0.026    0.011    0.717    0.535    1.990    0.871    0.196
      3    0.106    0.338    0.004    0.038    0.011    0.698    0.527    2.622    0.875    0.206
      4    0.107    0.421    0.103    0.077    0.018    0.678    0.512    3.167    0.880    0.214
:::::::::
      1    0.248    0.166    0.006    0.016    0.011    0.752    0.528    1.382    0.885    0.181
      2    0.247    0.251    0.056    0.020    0.011    0.704    0.531    1.963    0.858    0.193
      3    0.250    0.339    0.069    0.031    0.011    0.669    0.517    2.520    0.845    0.203
      4    0.250    0.422   -0.009    0.060    0.012    0.653    0.496    3.062    0.855    0.210
:::::::::
      1    0.410    0.165    0.029    0.020    0.011    0.757    0.523    1.379    0.888    0.175
      2    0.413    0.252    0.138    0.025    0.013    0.688    0.510    1.924    0.842    0.186
      3    0.411    0.338    0.074    0.034    0.013    0.658    0.493    2.475    0.835    0.195
      4    0.406    0.422    0.056    0.066    0.018    0.658    0.471    3.085    0.862    0.202
:::::::::
      1    0.574    0.168    0.032    0.029    0.010    0.745    0.521    1.384    0.878    0.166
      2    0.572    0.251    0.116    0.030    0.012    0.697    0.503    1.942    0.853    0.176
      3    0.566    0.336    0.075    0.046    0.014    0.698    0.483    2.609    0.877    0.185
      4    0.557    0.416    0.053    0.109    0.017    0.714    0.460    3.304    0.917    0.192
:::::::::
      1    0.733    0.166    0.164    0.041    0.018    0.754    0.521    1.390    0.887    0.154
      2    0.725    0.247    0.027    0.053    0.019    0.737    0.500    2.024    0.892    0.164
      3    0.716    0.331    0.108    0.109    0.013    0.759    0.476    2.800    0.935    0.173
:::::::::
      1    0.890    0.160   -0.135    0.084    0.019    0.779    0.495    1.388    0.909    0.140
      2    0.879    0.240   -0.030    0.165    0.023    0.786    0.463    2.101    0.936    0.149
\end{verbatim}
\eject

\section{Table 7: $A_{UL}^{\sin 2\phi_h}  vs.\ P_T$}
\begin{verbatim}
%% CLAS eg1-dvcs SIDIS azimuthal asymmetries for neutral pions
%% The table quotes the average value within a bin.
%% P_T is the momentum of the pi0 transverse to the momentum transfer direction (in GeV).
%% x is the Bjorken scaling variable.
%% A_UL^sin2 is the measured sine-2phi moment of the target spin asymmetry for longitudinally polarized protons.
%% Delta_stat is the statistical uncertainty on the moment.
%% Delta_sys is the systematic uncertainty on the moment.
%% y is the fractional energy transfer.
%% z is the fractional energy of the pi0.
%% Q^2 is the 4-momentum transfer squared.
%% depol is the depolarization factor, which accounts for the mismatch between the beam and momentum transfer directions.
%% dilut is the applied dilution factor.

  P_T-bin#  P_T       x   A_UL^sin2 Delta_stat Delta_sys  y       z       Q^2     depol     dilut
      1    0.106    0.168    0.017    0.024    0.010    0.748    0.534    1.394    0.882    0.185
      2    0.248    0.166    0.029    0.017    0.011    0.752    0.528    1.382    0.885    0.181
      3    0.410    0.165    0.053    0.021    0.015    0.757    0.523    1.379    0.888    0.175
      4    0.574    0.168    0.041    0.026    0.016    0.745    0.521    1.384    0.878    0.166
      5    0.733    0.166   -0.058    0.040    0.015    0.754    0.521    1.390    0.887    0.154
      6    0.890    0.160   -0.040    0.090    0.026    0.779    0.495    1.388    0.909    0.140
:::::::::
      1    0.106    0.250   -0.024    0.025    0.012    0.717    0.535    1.990    0.871    0.196
      2    0.247    0.251    0.012    0.020    0.012    0.704    0.531    1.963    0.858    0.193
      3    0.413    0.252   -0.046    0.023    0.013    0.688    0.510    1.924    0.842    0.186
      4    0.572    0.251   -0.011    0.027    0.016    0.697    0.503    1.942    0.853    0.176
      5    0.725    0.247   -0.120    0.053    0.017    0.737    0.500    2.024    0.892    0.164
      6    0.879    0.240    0.124    0.181    0.056    0.786    0.463    2.101    0.936    0.149
:::::::::
      1    0.106    0.338    0.014    0.037    0.012    0.698    0.527    2.622    0.875    0.206
      2    0.250    0.339   -0.011    0.028    0.012    0.669    0.517    2.520    0.845    0.203
      3    0.411    0.338   -0.024    0.030    0.011    0.658    0.493    2.475    0.835    0.195
      4    0.566    0.336   -0.046    0.040    0.015    0.698    0.483    2.609    0.877    0.185
      5    0.716    0.331    0.002    0.104    0.013    0.759    0.476    2.800    0.935    0.173
:::::::::
      1    0.107    0.421   -0.144    0.071    0.013    0.678    0.512    3.167    0.880    0.214
      2    0.250    0.422   -0.028    0.050    0.011    0.653    0.496    3.062    0.855    0.210
      3    0.406    0.422    0.090    0.054    0.018    0.658    0.471    3.085    0.862    0.202
      4    0.557    0.416    0.119    0.092    0.031    0.714    0.460    3.304    0.917    0.192
\end{verbatim}
\eject

\section{Table 8: $A_{UL}^{\sin 2\phi_h}  vs.\ x$}
\begin{verbatim}
%% CLAS eg1-dvcs SIDIS azimuthal asymmetries for neutral pions
%% The table quotes the average value within a bin.
%% P_T is the momentum of the pi0 transverse to the momentum transfer direction (in GeV).
%% x is the Bjorken scaling variable.
%% A_UL^sin2 is the measured sine-2phi moment of the target spin asymmetry for longitudinally polarized protons.
%% Delta_stat is the statistical uncertainty on the moment.
%% Delta_sys is the systematic uncertainty on the moment.
%% y is the fractional energy transfer.
%% z is the fractional energy of the pi0.
%% Q^2 is the 4-momentum transfer squared.
%% depol is the depolarization factor, which accounts for the mismatch between the beam and momentum transfer directions.
%% dilut is the applied dilution factor.

  x-bin#  P_T       x     A_UL^sin2  Delta_stat Delta_sys  y      z       Q^2     depol     dilut
      1    0.106    0.168    0.017    0.024    0.010    0.748    0.534    1.394    0.882    0.185
      2    0.106    0.250   -0.024    0.025    0.012    0.717    0.535    1.990    0.871    0.196
      3    0.106    0.338    0.014    0.037    0.012    0.698    0.527    2.622    0.875    0.206
      4    0.107    0.421   -0.144    0.071    0.013    0.678    0.512    3.167    0.880    0.214
:::::::::
      1    0.248    0.166    0.029    0.017    0.011    0.752    0.528    1.382    0.885    0.181
      2    0.247    0.251    0.012    0.020    0.012    0.704    0.531    1.963    0.858    0.193
      3    0.250    0.339   -0.011    0.028    0.012    0.669    0.517    2.520    0.845    0.203
      4    0.250    0.422   -0.028    0.050    0.011    0.653    0.496    3.062    0.855    0.210
:::::::::
      1    0.410    0.165    0.053    0.021    0.015    0.757    0.523    1.379    0.888    0.175
      2    0.413    0.252   -0.046    0.023    0.013    0.688    0.510    1.924    0.842    0.186
      3    0.411    0.338   -0.024    0.030    0.011    0.658    0.493    2.475    0.835    0.195
      4    0.406    0.422    0.090    0.054    0.018    0.658    0.471    3.085    0.862    0.202
:::::::::
      1    0.574    0.168    0.041    0.026    0.016    0.745    0.521    1.384    0.878    0.166
      2    0.572    0.251   -0.011    0.027    0.016    0.697    0.503    1.942    0.853    0.176
      3    0.566    0.336   -0.046    0.040    0.015    0.698    0.483    2.609    0.877    0.185
      4    0.557    0.416    0.119    0.092    0.031    0.714    0.460    3.304    0.917    0.192
:::::::::
      1    0.733    0.166   -0.058    0.040    0.015    0.754    0.521    1.390    0.887    0.154
      2    0.725    0.247   -0.120    0.053    0.017    0.737    0.500    2.024    0.892    0.164
      3    0.716    0.331    0.002    0.104    0.013    0.759    0.476    2.800    0.935    0.173
:::::::::
      1    0.890    0.160   -0.040    0.090    0.026    0.779    0.495    1.388    0.909    0.140
      2    0.879    0.240    0.124    0.181    0.056    0.786    0.463    2.101    0.936    0.149
\end{verbatim}



\end{document}